\documentclass[runningheads]{llncs}
\usepackage{amsmath,amsfonts}
\usepackage{graphicx}
\usepackage{multirow,multicol,booktabs}
\usepackage[dvipsnames]{xcolor}
\usepackage[numbers,sort&compress]{natbib}
\usepackage{personalcommand}
\usepackage{physics} 
\PassOptionsToPackage{hyphens}{url}\usepackage[hidelinks]{hyperref}\usepackage{cleveref}
\usepackage{colortbl}
\usepackage{tcolorbox}
\usepackage{subcaption}
\usepackage[loose]{units}

\newcommand{\expComp}[1]{\textbf{#1} UML Components\xspace}
\newcommand{\expNode}[1]{\textbf{#1} UML Nodes\xspace}
\newcommand{\expUC}[1]{\textbf{#1} UML Use Cases\xspace}

\newcommand{\SoBigDataITHack}{European Union - NextGenerationEU - National Recovery and Resilience Plan (Piano Nazionale di Ripresa e Resilienza, PNRR) - Project: “SoBigData.it - Strengthening the Italian RI for Social Mining and Big Data Analytics” - Prot. IR0000013 - Avviso n. 3264 del 28/12/2021\xspace}

\begin{document}

\title{Performance of Genetic Algorithms in the Context of Software Model Refactoring}
\author{Vittorio Cortellessa\inst{1}\orcidID{0000-0002-4507-464X}
Daniele {Di Pompeo}\inst{1}\orcidID{0000-0003-2041-7375} \and
Michele Tucci\inst{1}\orcidID{0000-0002-0329-1101}}
\authorrunning{Cortellessa et al.}
\institute{University of L'Aquila, L'Aquila, Italy\\ 
\email{\{vittorio.cortellessa, daniele.dipompeo, michele.tucci\}@univaq.it}}
\maketitle              \begin{abstract}
Software systems continuously evolve due to new functionalities, requirements, or maintenance activities.
In the context of software evolution, software refactoring has gained a strategic relevance. 
The space of possible software refactoring is usually very large, as it is given by the combinations of different refactoring actions that can produce software system alternatives. 
Multi-objective algorithms have shown the ability to discover alternatives by pursuing different objectives simultaneously. 
Performance of such algorithms in the context of software model refactoring is of paramount importance. Therefore, in this paper, we conduct a performance analysis of three genetic algorithms to compare them in terms of performance and quality of solutions. 
Our results show that there are significant differences in performance among the algorithms (\eg \pesa seems to be the fastest one, while \nsga shows the least memory usage).

\end{abstract}

\section{Introduction}\label{sec:intro}

Multi-objective optimization techniques proved to be effective in tackling many model-driven software development problems~\cite{Ramirez:2018uz,Mariani:2017jd,Ouni:2017db,Kessentini:2012cb}.
Such problems usually involve a number of quantifiable metrics that can be used as objectives to drive the optimization.
Problems related to non-functional aspects undoubtedly fit into this category, as confirmed by the vast literature in this domain~\cite{Aleti:2013gp,Koziolek:2011cg,DBLP:conf/mompes/AletiBGM09}. 
Most approaches are based on evolutionary algorithms~\cite{DBLP:journals/csur/BlumR03}, which allow exploring the solution space by combining solutions.

The improvement of software models quality through refactoring is a kind of task that can be carried out by multi-objective optimization.
However, multi-objective algorithms demand a lot of hardware resources (\eg time, and memory allocation) to search the solution space and generate a (near-)optimal Pareto frontier.
Therefore, the actual performance of multi-objective algorithms in software model refactoring is of paramount importance, especially if the goal is to integrate them into the design and evolution phases of software development.

For this reason, in this paper, we compare the performance in terms of execution time, memory allocation, and quality of Pareto frontiers of the \nsga, \spea, and \pesa multi-objective algorithms within the context of software model refactoring.
We have selected \nsga due to its extensive use in the context of software refactoring, \spea because it has already been compared with \nsga in other domains~\cite{CORTELLESSA2021106568,king2010comparison,GADHVI2016361}, and \pesa because it uses a different technique (\ie hyper-grid crowding degree operator) to search the solution space.

We have evaluated the performance of each algorithm by using a reference case study presented in~\cite{DBLP:conf/staf/Pompeo0CE19}. 
To achieve this, we have executed \emph{30} independent runs, as suggested in \citep{DBLP:conf/icse/ArcuriB11}, for each algorithm by varying the number of iterations for each run, and we have collected execution time and memory usage. 
We provide a replication package of the experimentation presented in this study.\footnote{Replication package: \url{https://github.com/danieledipompeo/replication-package__Perf-Comp-GA-4-Multi-Obj-SW-Model-Ref}}

We aim at answering the following research questions:
\begin{itemize}
    \item \textbf{$RQ_1$}: \emph{How do NSGA-II, SPEA2, and PESA2 compare in terms of execution time?}
    \item \textbf{$RQ_2$}: \emph{How do NSGA-II, SPEA2, and PESA2 compare in terms of memory usage?}
    \item \textbf{$RQ_3$}: \emph{How do NSGA-II, SPEA2, and PESA2 compare in terms of multi-objectindicatorsive optimization indicators?}
\end{itemize}
Our experimentation showed that \pesa is the algorithm whose executions last quite less than the \nsga and \spea ones. Furthermore, \pesa generates Pareto frontiers that showed better solutions in terms of reliability and performance. \nsga, instead, consumed less memory than \spea, and \pesa. However, it generated less densely populated Pareto frontiers. Finally, \spea showed worse performance and Pareto frontiers than \pesa, and \nsga.

The remaining of the paper is structured as follows:
\Cref{sec:related} reports related work; \Cref{sec:background:algos} introduces the algorithms subject of the study; \Cref{sec:case-study} briefly introduces the case studies; \Cref{sec:results} discusses results and findings. \Cref{sec:takeaways} describes takeaways from the study; \Cref{sec:t2v} discussed threats to validity. \Cref{sec:conclusion} concludes the paper.
 \section{Related Work}\label{sec:related}

Genetic algorithms are exploited in different domains to identify alternatives of the initial problem that show at least one better attribute (\ie at least on objective). In particular, studies have analyzed the performance in building Pareto frontiers in heterogeneous domains, which span from automotive problems to economic ones~\cite{1554690,king2010comparison,zhao2016improved,DBLP:journals/infsof/CortellessaPST23}. In this paper, instead, we analysed performance in terms of hardware consumption needed to search the solution space for software model refactoring.
In the context of software architecture, studies have investigated how multi-objective optimization can improve quality of software architectures.

For example, \citet{CORTELLESSA2021106568} studied the sensitivity of multi-objective software architecture refactoring to configuration characteristics. 
They compared two genetic algorithms in terms of Pareto frontiers quality dealing with architectures defined in \AE milia, which is a perfor\-mance-oriented Architecture Descrption Language (ADL). 
In this paper, we propose a performance comparison between \nsga, \spea, and \pesa to identi\-fy which algorithm needs less resources to search the solution space.

\citet{DBLP:conf/mompes/AletiBGM09} have presented an approach for modeling and analyzing Architecture Analysis and Design Language (AADL) architectures~\cite{DBLP:books/daglib/0030032}. 
They have also introduced a tool aimed at optimizing different quality attributes while varying the architecture deployment and the component redundancy. 
Instead, our work relies on UML models and offers more complex refactoring actions as well as different target attributes for the fitness function. 
Besides, we investigate the role of performance antipatterns in the context of multi-objective software architecture refactoring optimization.

\citet{DBLP:conf/wosp/MenasceEGMS10} have presented a framework for architectural design and quality optimization, where architectural patterns are used to support the searching process (\eg load balancing, fault tolerance). 
Two limitations affects the approach: the architecture has to be designed in a tool-related notation and not in a standard modelling language (as we do in this paper), and it uses equation-based analytical models for performance indices that might be too simple to capture architectural details and resource contention. 
We overcome the possible the \citeauthor{DBLP:conf/wosp/MenasceEGMS10} limitation by employing Layred Queueing Network (LQN) models to estimate performance indices.

\citet{Martens:2010bn} have presented PerOpteryx, a performance-oriented multi-objective optimization problem. 
In PerOpteryx the optimization process is guided by tactics referring to component reallocation, faster hardware, and more hardware, which do not represent structured refactoring actions, as we employ in our refactoring engine. 
Moreover, PerOpteryx supports architectures specified in Palladio Component Model (PCM)~\cite{Becker:2009cl} and produces, through model transformation, a LQN for of performance analysis. 

\citeauthor{10.1145/3132498.3132509} have presented SQuAT~\cite{10.1145/3132498.3132509}, an extensible platform aimed at including flexibility in the definition of an architecture optimization problem. 
SQuAT supports models conforming to PCM language, exploits LQN for performance evaluation, and PerOpteryx tactics for architecture.

A recent work compares the ability of two different multi-objective optimization approaches to improve non-functional attributes~\citep{NI2021106565}, where randomized search rules have been applied to improve the software model. 
The study of \citet{NI2021106565} is based on a specific modelling notation (\ie PCM) and it has implicitly shown that the multi-objective optimization problem at model level is still an open challenge. 
They applied architectural tactics, which in general do not represent structured refactoring actions, to find optimal solutions. 
Conversely, we applied refactoring actions that change the structure of the initial model by preserving the original behavior. 
Another difference is the modelling notation, as we use UML with the goal of experimenting on a standard notation instead of a custom DSL.

\section{Algorithms}\label{sec:background:algos}

\paragraph{NSGA-II}\label{sec:background:nsga} 
The Non-dominated Sorting Algorithm II (\nsga), introduced by \citet{Deb:2002ut}, is widely used in the software engineering community due to its good performance in generating Pareto frontiers. 
The algorithm, randomly generates the initial population ${P_0}$, shuffles it and applies the \emph{Crossover} operator with probability $P_{crossover}$, and the \emph{Mutation} operator with probability $P_{Mutation}$ to generate the $Q_t$ offspring. 
Thus, the obtained $R_t=P_t+Q_t$ mating pool is sorted by the \emph{Non-dominated sorting} operator, which lists Pareto frontiers with respect to considered objectives. 
Finally, a \emph{Crowding distance} is computed and a new family (\ie $P_{t+1}$) is provided to the next step by cutting the worse half off.

\paragraph{SPEA2}\label{sec:background:spea} 
Strength Pareto Evolutionary Algorithm 2 (\spea) has been introduced by \citet{zitzler2001spea2}. 
Differently from \nsga, \spea does not employ a non-dominated sorting process to generate Pareto frontiers.

\spea randomly generates an initial population $P_0$ and an empty archive $\bar{P}_0$ in which non-dominated individuals are copied at each iteration. 
For each iteration $t=0,1,\dots, T$, the fitness function values of individuals in $P_t$ and $\bar{P}_t$ are calculated.
Then non-dominated individuals of $P_t$ and $\bar{P}_t$ are copied to $\bar{P}_{t+1}$ by discarding dominated individuals or duplicates (with respect to the objective values). 
In case size of $\bar{P}_{t+1}$ exceeds $\bar{N}$, \ie the size of the initial population, the \emph{Truncation} operator drops exceeded individuals by preserving the characteristics of the frontier, using the \emph{k-$th$ nearest neighbor} knowledge. 
In case size of $\bar{P}_{t+1}$ is less than $\bar{N}$, dominated individuals from $P_t$ and $\bar{P}_{t}$ are used to fill $\bar{P}_{t+1}$.
The algorithm ends when a stopping criterion is met, \eg the iteration $t$ exceeds the maximum number of iterations $T$, and it generates the non-dominated set $A$ in output.

\paragraph{PESA2}\label{sec:background:pesa}
The Pareto Envelope-based Selection Algorithm 2 (\pesa) is a multi-objective algorithm, introduced by \citet{10.5555/2955239.2955289} that uses two sets of population, called internal (\emph{IP}) and external (\emph{EP}).The internal population is often smaller than the external one and it contains solution candidates to be included in the external population. 
Furthermore, the external population is generally called \emph{archive}. 
The selection process is driven by a hyper-grid crowding distance degree. 
The current set of \emph{IP} are incorporated into the \emph{EP} one by one if it is non-dominated within \emph{IP}, and if is not dominated by any current member of the \emph{EP}. 
Once a candidate has entered the \emph{EP}, members of the \emph{EP} which it dominated (if any) will be removed. 
If the addition of a candidate renders the \emph{EP} over-full, then an arbitrary chromosome which has the maximal squeeze factor in the population of \emph{EP} is removed. 
Also, the squeeze factor describes the total number of other chromosomes in the archive which inhabit the same box.
The \pesa crowding strategy works by forming an implicit hyper-grid which divides the solution space into hyper-boxes. 
Furthermore, each chromosome in the \emph{EP} is associated with a particular hyper-box in solution space. 
Then, the squeeze factor is assigned to each hyper-box, and it is used during the searching phase.

 \section{Case study}\label{sec:case-study}

In this section, we apply our approach to the Train Ticket Booking Service (\ttbs) case study~\cite{DBLP:conf/staf/Pompeo0CE19,DBLP:journals/tse/ZhouPXSJLD21}, and to the well-established model case study \ccm, whose UML model has been derived by the specification in~\cite{Herold2008}.

\subsubsection{Train Ticket Booking Service}

Train Ticket Booking Service (\ttbs) is a web-based booking application, whose architecture is based on the microservice paradigm. 
The system is made up of 40 microservices, and it provides different scenarios through users that can perform realistic operations, \eg book a ticket or watch trip information like intermediate stops. 
Our UML model of \ttbs is available online.\footnote{\url{https://github.com/SEALABQualityGroup/2022-ist-replication-package/tree/main/case-studies/train-ticket}}
The static view is made of \expComp{11}, where each component represents a microservice. In the deployment view, we consider \expNode{11}, each one representing a docker container.
We selected these three scenarios because they commonly represent performance-critical ones in a ticketing booking service.

\subsubsection{CoCOME}

\ccm describes a Trading System containing several stores. 
A store might have one or more cash desks for processing goodies. 
A cash desk is equipped with all the tools needed to serve  a customer (e.g., a Cash Box, Printer, Bar Code Scanner). 
\ccm describes 8 scenarios involving more than 20 components. 
From the \ccm original specification, we analyzed different operational profiles, \ie scenarios triggered by different actors (such as Customer, Cashier, StoreManager, StockManager),   and we excluded those related to marginal parts of the system, such as scenarios of the \emph{EnterpriseManager} actor.
Thus, we selected \expUC{3}, \expComp{13}, and \expNode{8} from the \ccm specification.

 \section{Results}\label{sec:results}

In this section, we compare execution times, memory consumption, and quality of Pareto frontiers across the considered algorithms and case studies.

\subsection{$RQ_1$: How do \nsga, \spea, and \pesa compare in terms of execution time?}\label{sec:results:time}
In order to answer to the $RQ_1$ we collected execution time of each algorithm 30 times. Based on the results of our experimentation, we can state that the \pesa algorithm showed the best execution time with respect to \nsga and \spea in both case studies. 
Also, it appears as complexity and size of the case study plays an important role in determining execution time and its variability across iterations.

\begin{figure}[htbp]
    \centering
    \begin{subfigure}{.49\linewidth}
        \includegraphics[width=\linewidth]{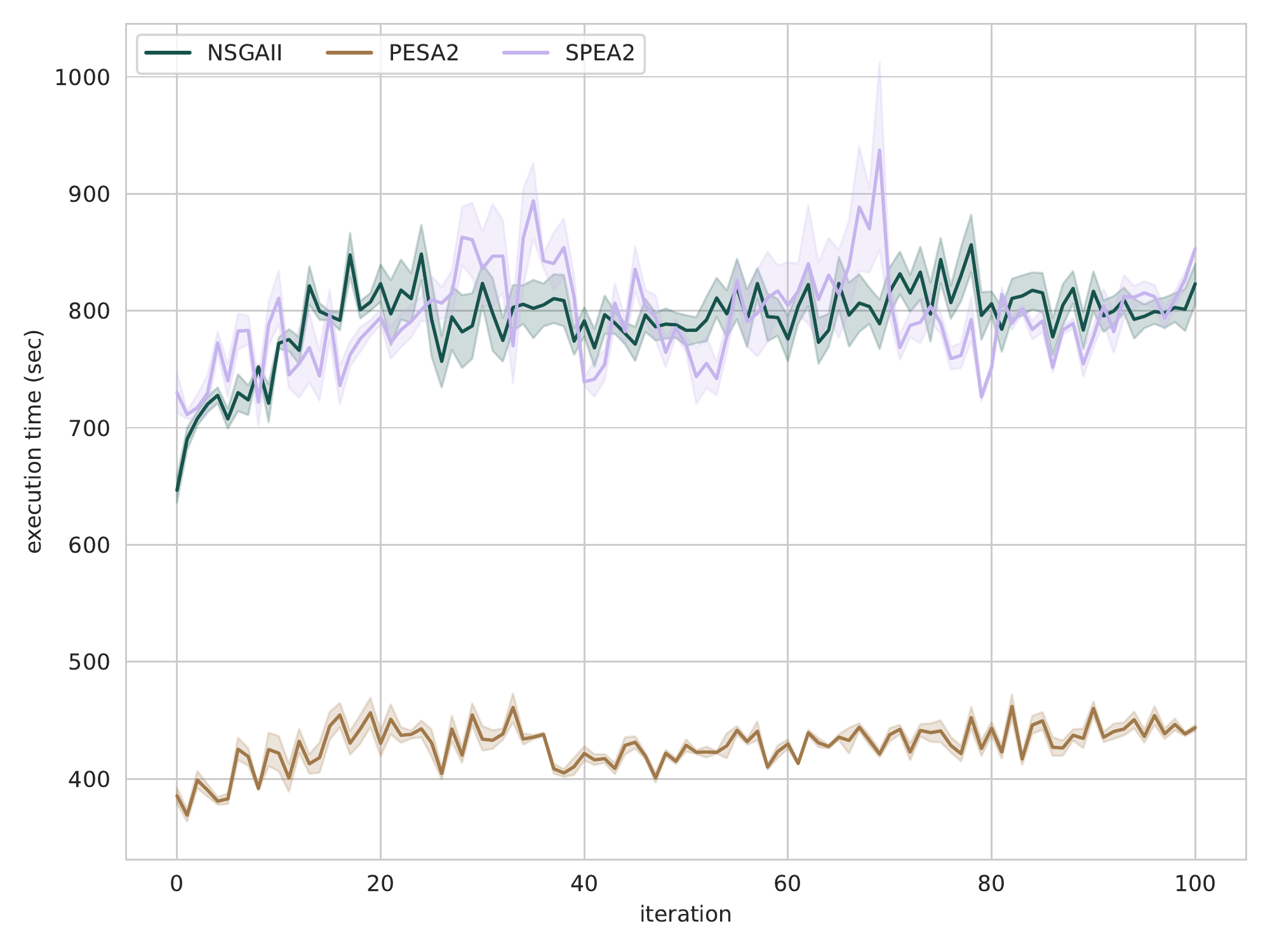}
        \caption{\ttbs}
        \label{fig:cmp_102_exectime_ttbs}
    \end{subfigure}
    \hfill \begin{subfigure}{.49\linewidth}
        \includegraphics[width=\linewidth]{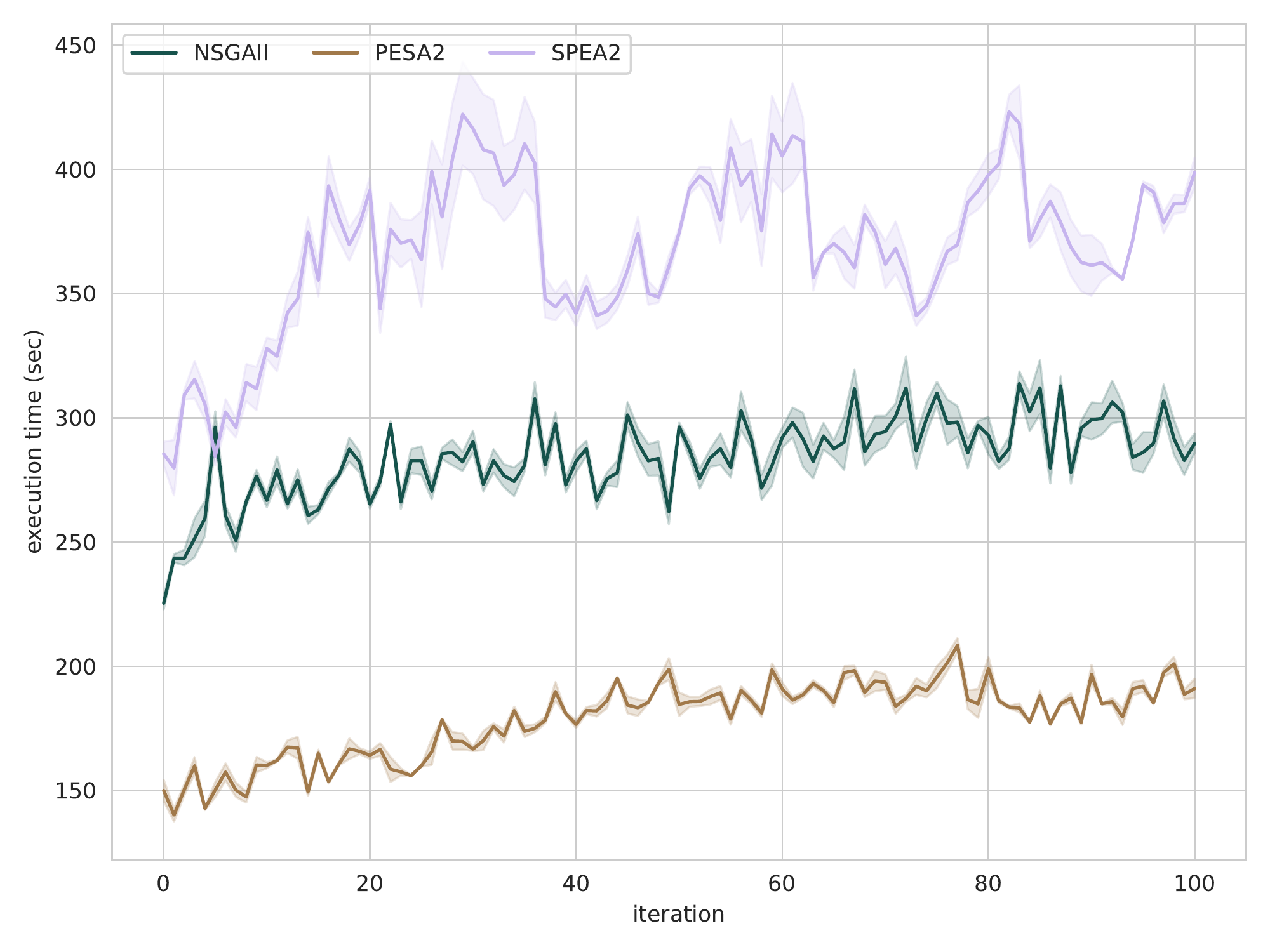}
        \caption{\ccm}
        \label{fig:cmp_102_exectime_ccm}
    \end{subfigure}
    \caption{Comparison of algorithms execution time.}
    \label{fig:cmp_102_exectime}
\end{figure}

\Cref{fig:cmp_102_exectime} compares \nsga, \spea, and \pesa in terms of their execution times for \ttbs and \ccm, respectively.
Darker lines report the mean over 30 runs for each iteration, while the bands represent 95\% confidence intervals for the mean, and are computed for the same runs.
Our results show substantial differences in the execution times of the algorithms, both on the same case study, and across them.

It is easy to notice that, regardless of the algorithm, the search is twice as fast in \ccm that it is in \ttbs, as it is obvious when observing the scale on the y-axis. 
\pesa is clearly the fastest algorithm in both cases (around \unit[400]{sec} in \ttbs, and \unit[180]{sec} in \ccm).
However, when it comes to comparing \nsga and \spea, their execution time, while consistently larger than \pesa, is almost on par in \ttbs, and noticeably apart in \ccm.
This suggests that the execution time might very well be dependent on the complexity and size of the specific case study.
For instance, it looks like the more complex the case study is, the slower \spea is.
Therefore, it appears evident that the search policy used by \spea, \ie the dominance operator, is slower than the crowing distance used by \nsga.
Moreover, the search policy employed by \pesa, \ie the hyper-grid crowding distance, seems to be faster than the ones used by \nsga and \spea, as it lasts half the time of the other two techniques.

Another interesting point, could be the stability of execution times, as it appears that the three algorithms exhibit different variability.
For instance, \pesa and \nsga showed a more stable execution time in both the case studies, while \spea showed a quite stable execution time with \ttbs, and a considerably larger variability with \ccm, with some abrupt changes.
This might be due to the usage of the archive for storing generated solutions.
When the case study is more complex, as it is the case for \ttbs, the usage of the archive seems to help find a Pareto frontier, while the usage of two archives with a less complex case study results in prolonged executions.
In fact, when a higher number of different solutions are found, these slower executions may be caused by the fact that a higher number of comparisons are needed to fill the two archives.

\subsection{$RQ_2$: How do \nsga, \spea, and \pesa compare in terms of memory usage?}\label{sec:results:memory}
In order to answer to the $RQ_2$ we collected the memory allocation of each algorithm during the experiments by exploiting the Java API.
From our experimentation results, the \nsga algorithm shows the least memory consumption with respect to \pesa and \spea.
Our results also show that the memory usage is not strictly related to the complexity of the case study.

\begin{figure}[htbp]
    \centering
    \begin{subfigure}{.49\linewidth}
    \includegraphics[width=\columnwidth]{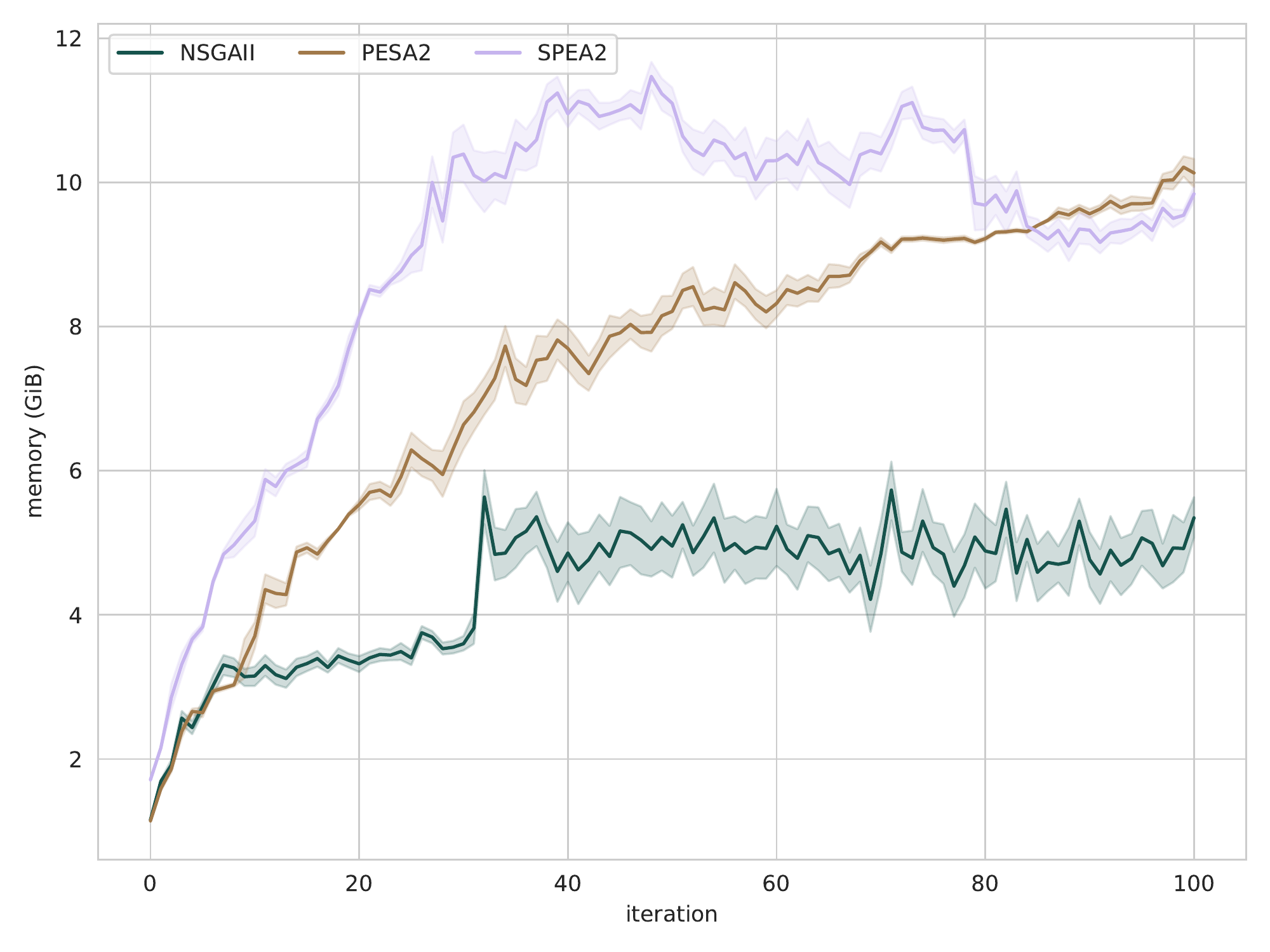}
    \caption{\ttbs}
    \label{fig:cmp_102_memory_ttbs}
\end{subfigure}
\hfill \begin{subfigure}{.49\linewidth}
    \includegraphics[width=\columnwidth]{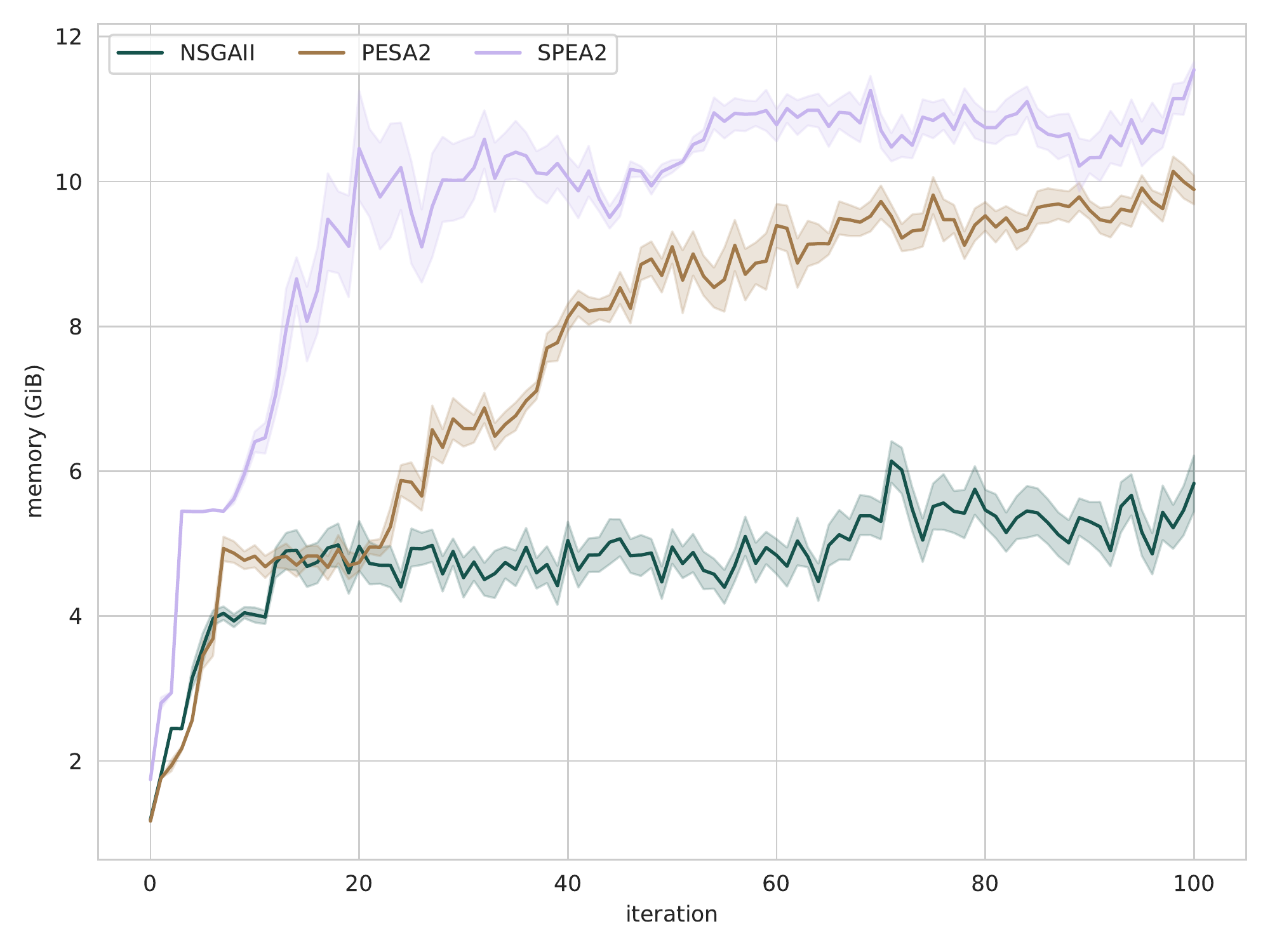}
    \caption{\ccm}
    \label{fig:cmp_102_memory_ccm}
\end{subfigure}
    \caption{Comparison of algorithms memory usage.}
    \label{fig:cmp_102_memory}
\end{figure}

\Cref{fig:cmp_102_memory} shows the memory allocation of the three algorithms. 
\nsga, \spea, and \pesa occupy the same quantity of memory showing an increase trend of the memory usage around the 20 iterations, then \nsga becomes almost flat.
Moreover, \spea shows a steep memory usage, and it occupies all the available memory after 40 iterations, while \pesa showed a smooth but linear increase of the memory, and it filled the available memory after 80 iteration.

Undoubtedly, the \nsga search policy is the least memory demanding among the three analyzed in our study, and it requires around \unit[5]{GiB} when it stabilizes.
\spea and \pesa, on the other hand, occupy almost all the available memory (\ie \unit[12]{GiB}).

\spea shows a different behavior in the two case studies, in our results.
In the case of \ccm, we can see an almost flat memory consumption around the \unit[12]{Gib} after 20 iteration, while in \ttbs we can observe a reduction of the memory allocation after 80 iterations.
Combining the latter with the overall quality of the generated Pareto fronts (see \Cref{sec:results:pareto}), we can assume that \spea cannot find better solutions after 80 iterations, thus any new solution was probably already stored in the two archives.

Finally, \pesa showed an interesting trend, as it allocated more memory almost linearly.
This might be due to the search policy of splitting the solution space in hyper-grids that will require to store new solutions when other locations of the solution space will be investigated by longer iterations. 
Therefore, we can expect that \pesa will likely exceed the \unit[12]{GiB} with longer iterations.

\subsection{$RQ_3$: How do \nsga, \spea, and \pesa compare in terms of multi-objective optimization indicators?}\label{sec:results:pareto}
In order to answer to the $RQ_3$ we graphically compare properties of the Pareto frontiers computed by each algorithm, and we use well-known indicators to estimate the quality of Pareto frontiers. 
From our results, the \pesa algorithm is the best to search the solution space, with solutions closest to the reference Pareto in \ttbs and \ccm. 
Also, \nsga generates solutions with the highest variability in both case studies.
Finally, \spea did not show any quality indicators with higher quality.

\begin{figure}
    \centering
    \begin{subfigure}{.49\linewidth}
    \includegraphics[width=\columnwidth]{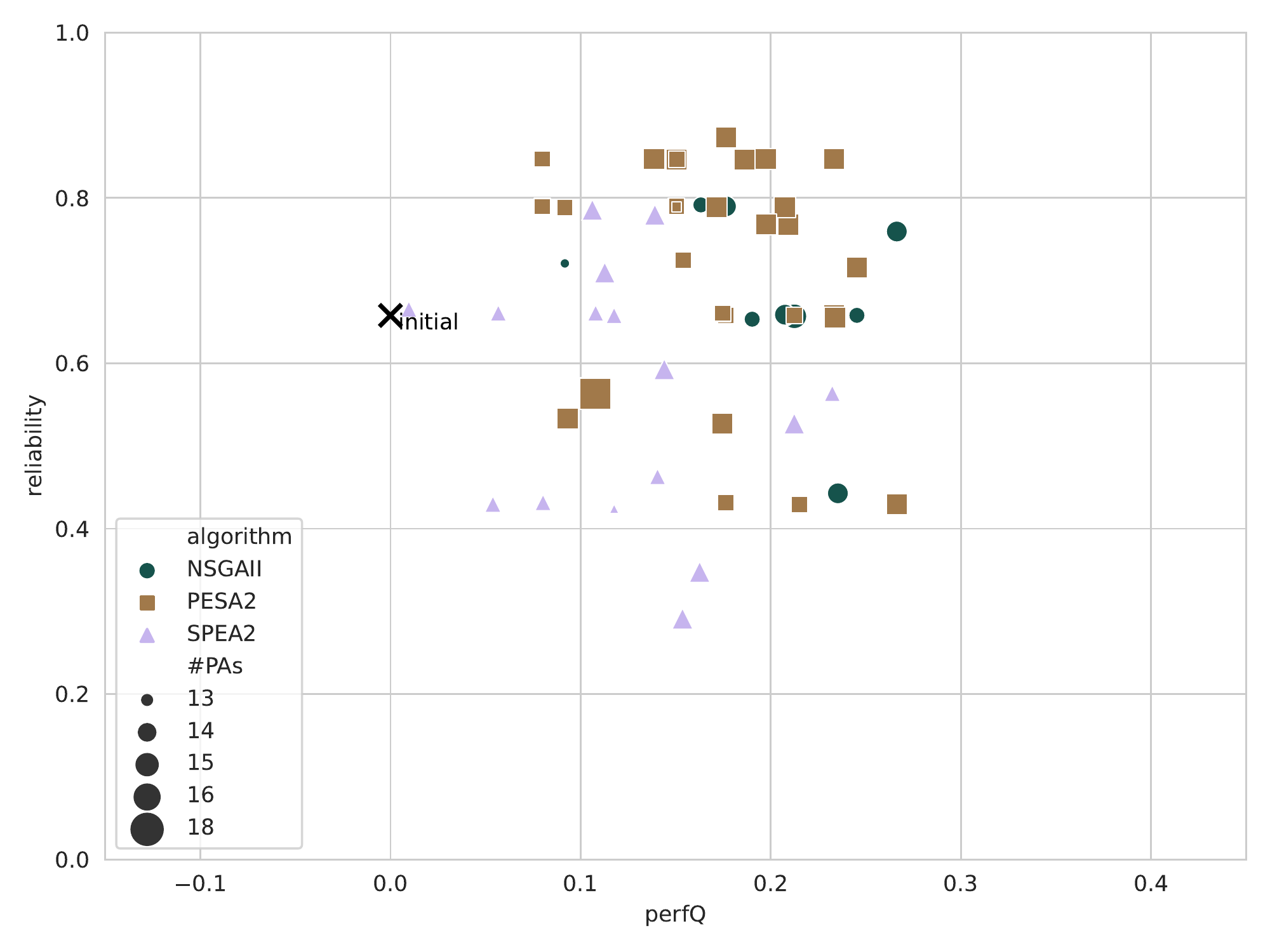}
    \caption{\ttbs.}
    \label{fig:cmp_reference_102_ttbs}
\end{subfigure}
\hfill \begin{subfigure}{.49\linewidth}
\includegraphics[width=\columnwidth]{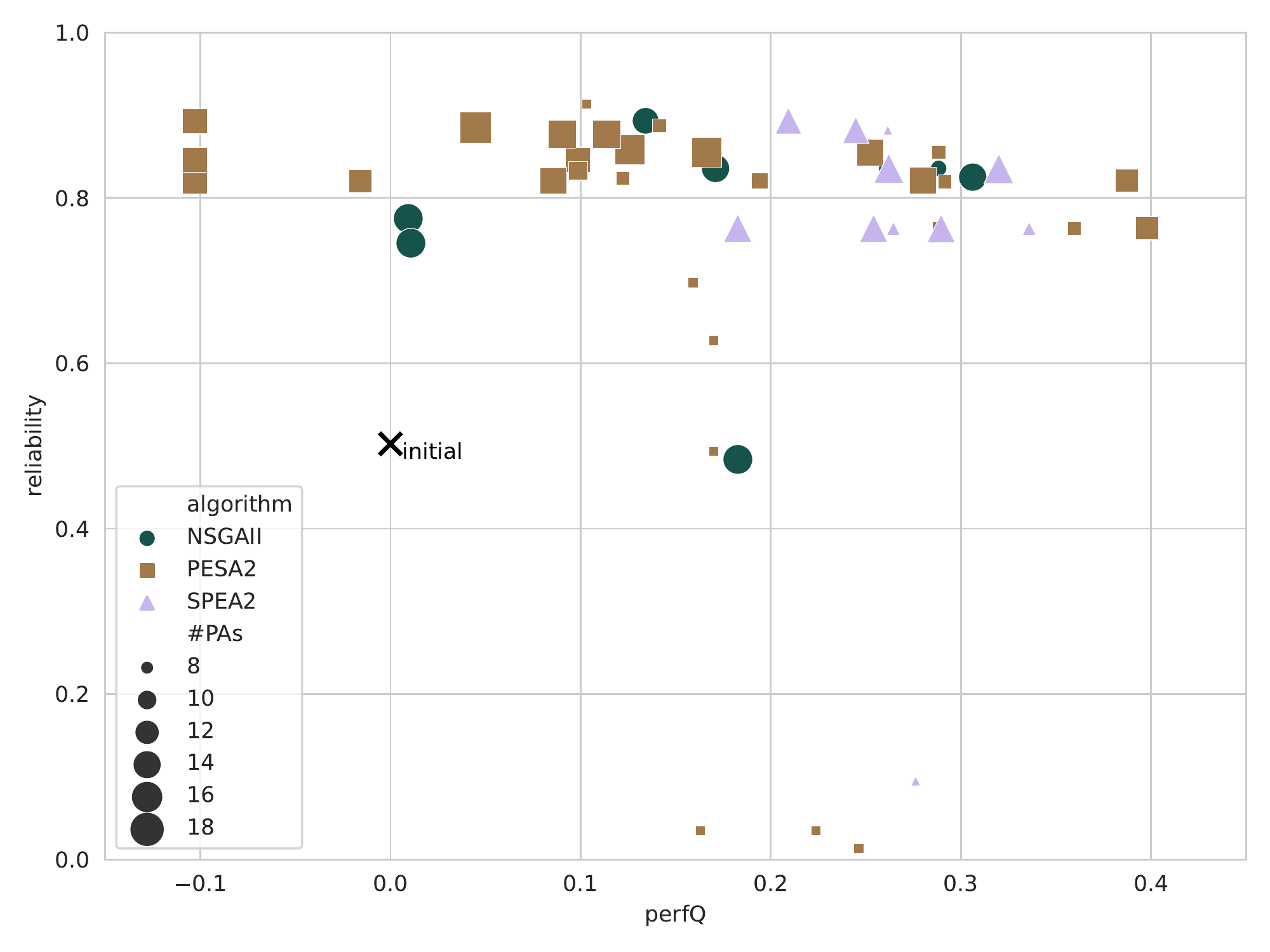}
    \caption{\ccm.}
    \label{fig:cmp_reference_102_ccm}
\end{subfigure}
    \caption{Comparison of reference Paretos.}
    \label{fig:cmp_reference_102}
\end{figure}

The overall quality of computed Pareto frontiers (\computedP) is one of the most critical parameters to consider when comparing genetic algorithms.
\Cref{fig:cmp_reference_102} depicts the \computedP generated by the three genetic algorithms for \ttbs and \ccm, where, in each plot, the top right quadrant is the optimal location for the optimization.
Furthermore, we measure the quality of \computedP through the quality indicators listed in \Cref{tab:qindicators}.

From \Cref{fig:cmp_reference_102_ttbs,fig:cmp_reference_102_ccm}, we can clearly deduce that none of the subject algorithms shows the ability of finding solutions towards the top right quadrant in both the case studies.
In fact, we can see that the solutions are organized in a vertical cluster in \Cref{fig:cmp_reference_102_ttbs}, and in a horizontal one in \Cref{fig:cmp_reference_102_ccm}.
Also, it appears that the optimization process selects similar refactoring actions, therefore generating almost identical solutions within the frontiers.

Furthermore, we can observe a different behavior of each algorithm in \ttbs, and \ccm. 
For example, \pesa found the best solutions for \ccm, in terms of \reliability and \perfq (\eg see the rightmost squares in \Cref{fig:cmp_reference_102_ccm}), while this is not the case for \ttbs, where, instead, \pesa found the best solution in terms of \perfq, with worse \reliability than the initial solution (see the square near the point (0.3, 0.4) in \Cref{fig:cmp_reference_102_ttbs}).

Besides the graphical analysis, we performed a study of the quality of \computedP in both case studies, by exploiting established quality indicators for multi-objective optimization.
It is important to recall that an indicator estimates the quality of a specific property of \computedP with respect to the reference Pareto frontier (\referenceP). 
Since \referenceP has not yet been defined for our case studies, we estimated the \referenceP as the set of non-dominated solutions produced by any algorithm.
In particular, we computed the \emph{Hypervolume (HV)}~\citep{DBLP:conf/ppsn/ZitzlerT98}, \emph{Inverted Generational Distance + (IGD+)}~\cite{DBLP:conf/gecco/LopezC18}, \emph{GSPREAD}~\cite{durillo2011jmetal}, and \emph{Epsilon (EP)}~\cite{durillo2011jmetal} quality indicators. 
We listed quality indicators \emph{(Q Ind)} in \Cref{tab:qindicators}, where the up arrow ($\uparrow$) means the indicator is to be maximized, and the down arrow ($\downarrow$) means the indicator is to be minimized.

From our experimental results, we see that \pesa produced the highest value of Hypervolume, thus proving that the algorithm covered the solution space better than \nsga and \spea. 
Also, \pesa showed the best value of IGD+, meaning that solutions belonging to the \computedP are closer to the \referenceP.
\nsga produced the best value of generalized spread (GSPREAD), thus indicating that the solutions in \nsga Pareto frontiers are more different from each other. 
Finally, our results prove that \spea computes quality indicators with good quality only for \ccm.
Therefore, it seems that \spea is able to find good \computedP when case studies with lower complexity.

\begin{table*}
    \centering
    \begin{tabular}{lcrrrrrr}
        \toprule
        \multicolumn{1}{c}{\multirow{2}{*}{Q Ind}} & \multicolumn{1}{c}{\multirow{2}{*}{\# iter}} & \multicolumn{2}{c}{\nsga} & \multicolumn{2}{c}{\pesa} & \multicolumn{2}{c}{\spea} \\
        \cmidrule(lr){3-4}
        \cmidrule(lr){5-6}
        \cmidrule(lr){7-8}
	{} & {} & \ttbs & \ccm & \ttbs & \ccm & \ttbs & \ccm \\
        
        \midrule
\multirow{1}{*}{HV~($\uparrow$)}        & 102  & 0.22433 &  0.07022  & 0.50909 & 0.44431 & 0.14467 & 0.36521 \\ \multirow{1}{*}{IGD+~($\downarrow$)}    & 102  & 0.11221 &  0.06005  & 0.04683 & 0.04046 & 0.10270 & 0.06620 \\\multirow{1}{*}{GSPREAD~($\downarrow$)} & 102  & 0.16013 &  0.12675  & 0.38391 & 0.52451 & 0.39153 & 0.33592 \\\multirow{1}{*}{EP~($\downarrow$)}      & 102  & 0.33333 &  0.20339  & 0.20000 & 0.10000 & 0.50000 & 0.36191 \\\bottomrule
    \end{tabular}
    \caption{Quality indicators to establish the overall quality of Pareto frontiers.}
    \label{tab:qindicators}
\end{table*}
  \section{Lesson Learned}\label{sec:takeaways}

Genetic algorithms have proved to help optimize quantifiable metrics, such as performance and reliability.
Their ability to search for optimal solutions is influenced by several configuration parameters.
In our experience, we noticed that each configuration parameter has a different impact on the overall performance, \eg the population size impacts the execution time and the memory usage.
A wider initial population size requires longer execution times to generate individuals, and it might produce stagnation during the optimization~\citep{DBLP:journals/ese/ArcuriF13} that, in turn, might hamper the quality of Pareto frontiers.
Furthermore, in model-based software refactoring, a wider initial population also implicates a higher memory consumption, because entire models need to be loaded in memory for the refactoring to be performed.
Hence, it is crucial to find the optimal trade-off between the configuration parameters and the quality of the Pareto frontiers.

Besides the initial population, crossover and mutation operators might impact the execution time.
For instance, a higher mutation probability will obviously produce more frequent mutations within the population.
The more mutations are produced, the higher the probability of having an invalid individual, thus requiring additional time to check for feasibility, repair or even change the individual entirely.
The crossover probability, instead, impacts the execution time since combinations of individuals are more frequent.
Furthermore, the crossover operator requires time to perform the combination and might also generate invalid individuals.
Therefore, using the right crossover and mutation probabilities is crucial for the time and quality of subsequent populations.
This is clearly an opportunity for further research on heuristic to estimate some configuration values, since it would be impractical to evaluate every parameter combination.
In future work, we plan to examine how different configurations affect the resulting quality and performance of different genetic algorithms.

We cannot guarantee that our analysis can be generalized in other domains or with other modeling notations.
However, in the context in which we performed our study, there is not an in-depth analysis of performance traits of genetic algorithms.
We believe this study might open a research direction on improving genetic algorithm performance for model-based refactoring optimization.
For example, in a recent work, \citet{SEAA2022} studied the possibility of reducing the search time by limiting it with a budget.
Also, knowing how algorithms compare in terms of performance might open to even using them in an interactive optimization process, where the designer could be involved in a step-by-step decision process aided by the automation provided by the algorithms, but bounded in time.
In such a scenario, the designer could be at the core of the process, potentially making optimization trade-offs at each step. \section{Threats to validity}\label{sec:t2v}

Our results may be affected by threats related to the specific versions of Java and the JMetal library.
We mitigated these threats by using the same configuration for the Java Virtual Machine and the same JMetal version in each run. 
In particular, we used the OpenJDK 11 with default configuration, and we built the implementation on JMetal v5.10.

Multiple factors may influence the performance measurements gathered during the experiments and, therefore, could undermine our conclusions.
However, we mitigated external influences by disabling any other software running on the same server, and we repeated each experiment 30 times as suggested in~\citep{DBLP:conf/icse/ArcuriB11}.

Also, the study we conducted in this paper may be affected, as for any performance analysis experiment, by the input data. 
Although we use two case studies, our deductions may change if other case studies, \ie different software models, are employed as inputs to the optimization problem. 
However, we considered two models presented in~\cite{DBLP:conf/staf/Pompeo0CE19,Herold2008} that has been already used in other performance analysis problems~\cite{CORTELLESSA2021111084,SEAA2021,SEAA2022}.
To the best of our knowledge, there are no previous studies that analyzed and compared performance of multi-objective algorithms in the context of software model refactoring, as we did in this study. 
Therefore, although the paper may suffer from this threat to conclusion validity, it represents a first investigation in this direction. 

The overall quality of Pareto frontiers generated by each algorithm has been estimated through well-known quality indicators. 
These indicators leverage the estimation by comparing a Pareto frontier to problem-specific reference points. 
Since in our experimentation these reference points are not yet available, we computed them as the non-dominated points within every Pareto frontier of each run of each algorithm. 
Therefore, the reference points might affect the overall quality computation, and we further investigate the usage of more appropriated reference points. 

Finally, our results may be affected by threats related to the configurations of the genetic algorithms. 
For example, the number of iterations can influence performance results. 
We cannot be sure to have effectively mitigated these threats because of the long execution time required to run each configuration. 
Such long execution times make trying many alternative configurations unfeasible. 
For this reason, we used a configuration in an attempt to detect performance flaws that may only manifest during longer executions. 
 \section{Conclusion}\label{sec:conclusion}

This study presented a performance comparison of three genetic algorithms, \ie \nsga, \spea, and \pesa. 
We selected those algorithms due to their wide usage in the software refactoring context and their search algorithm characteristics.

We compared the execution time, the memory allocation, and the quality of the produced Pareto frontiers. 
We collected performance metrics by using two case studies presented in~\citep{DBLP:conf/staf/Pompeo0CE19,Herold2008} by executing 30 different runs, and we compared the overall quality of Pareto frontiers through specific quality indicators, such as Hypervolume and IGD+.
Our analysis can summarize that \pesa is the fastest algorithm, and \nsga is the least memory-demanding algorithm. 
Finally, \spea has shown the worst memory usage as well as the worst execution time. 
We will further investigate memory consumption by employing a more sophisticated memory profiling, which might introduce an overhead within the measurements.

Concerning the overall quality of the produced Pareto frontiers, we found that \pesa produced the most densely populated Pareto frontiers, while \nsga generated the least densely populated frontiers. 
\pesa has also shown a linear memory consumption, thus we intend to further analyze the trend by exploring longer execution in terms of the number of iterations.

Furthermore, we intend to investigate if our findings can be generalized to other case studies, different algorithms, and different kinds of refactoring actions, as those aimed at other non-functional properties, such as availability~\cite{DBLP:conf/icsa/CortellessaET18}. 
\subsubsection*{Acknowledgements}
Daniele Di Pompeo and Michele Tucci are supported by \SoBigDataITHack.

\bibliographystyle{splncs04nat}
\bibliography{biblio}
\end{document}